\newif\ifsubmission
\submissiontrue 

\newif\iffullversion
\fullversiontrue

\newif\ifanonymous
\anonymousfalse 

\newif\ifchange
\changefalse 

\documentclass[journal]{IEEEtran}

\usepackage[utf8]{inputenc}
\usepackage{graphicx}
\usepackage[hyphens]{url}
\usepackage{hyperref}
\usepackage[inline]{enumitem}
\usepackage{float}
\usepackage{pifont}
\usepackage{amssymb}
\usepackage{tabularx}
\usepackage{listings}
\usepackage{booktabs}
\usepackage[nolist]{acronym}
\usepackage{multirow}

\usepackage{xcolor}
\definecolor{darkred}{rgb}{.6,0,0}
\definecolor{darkgreen}{rgb}{0,.4,0}
\definecolor{darkblue}{rgb}{0,0,.6}
\definecolor{darkpurp}{rgb}{.5,0,.5}
\definecolor{darkyellow}{rgb}{.78,.62,.02}

\usepackage[capitalise]{cleveref}

\usepackage{tikz}

\usetikzlibrary{automata,shapes,shadows,positioning,arrows,fadings,through,calc,trees,fit}

\pgfdeclarelayer{background}
\pgfsetlayers{background,main}

\tikzstyle{system}=[draw, fill=black!20, text width=3cm, text centered,
minimum height=1cm, drop shadow]
\tikzstyle{exploit}=[draw, fill=red!20, text width=2cm, text centered,
drop shadow]

\newfloat{lstfloat}{htbp}{lop}
\floatname{lstfloat}{Listing}

\newcolumntype{L}[1]{>{\hsize=#1\hsize\raggedright\arraybackslash}X}%
\newcolumntype{R}[1]{>{\hsize=#1\hsize\raggedleft\arraybackslash}X}%
\newcolumntype{C}[1]{>{\hsize=#1\hsize\centering\arraybackslash}X}%

\lstdefinelanguage{xml}
{
  morestring=[b]",
  morestring=[s]{>}{<},
  morecomment=[s]{<?}{?>},
  morekeywords={xmlns,version,type}
}

\lstdefinelanguage{cto}{
	keywords={namespace, asset, participant, abstract, extends, event,
	transaction},
}

\lstdefinelanguage{acl}{
	keywords={rule},
}

\lstdefinelanguage{javascript}{
  keywords={typeof, new, true, false, catch, function, return, null, catch, switch, var, if, in, while, do, else, case, break},
  ndkeywords={class, export, boolean, throw, implements, import, this},
  sensitive=false,
  comment=[l]{//},
  morecomment=[s]{/*}{*/},
  morestring=[b]',
  morestring=[b]"
}

\lstdefinelanguage{json}{
    literate=
     *{0}{{{\color{red!60!green}0}}}{1}
      {1}{{{\color{red!60!green}1}}}{1}
      {2}{{{\color{red!60!green}2}}}{1}
      {3}{{{\color{red!60!green}3}}}{1}
      {4}{{{\color{red!60!green}4}}}{1}
      {5}{{{\color{red!60!green}5}}}{1}
      {6}{{{\color{red!60!green}6}}}{1}
      {7}{{{\color{red!60!green}7}}}{1}
      {8}{{{\color{red!60!green}8}}}{1}
      {9}{{{\color{red!60!green}9}}}{1}
      {.}{{{\color{red!60!green}{.}}}}{1}
      {:}{{{\color{red!60!black}{:}}}}{1}
      {,}{{{\color{red!60!black}{,}}}}{1}
      {\{}{{{\color{red!60!black}{\{}}}}{1}
      {\}}{{{\color{red!60!black}{\}}}}}{1}
      {[}{{{\color{red!60!black}{[}}}}{1}
      {]}{{{\color{red!60!black}{]}}}}{1},
}

\newcommand{\eg}{\textit{e.g.}}
\newcommand{\ie}{\textit{i.e.}}
\newcommand{\etal}[1]{#1~\emph{et al.}}
\newcommand{\xmark}{\ding{55}}%
\newcommand{\extrust}{\textsc{ExTRUST}}
\newcommand{\bcextrust}{BC-based \extrust}
\newcommand{\mpcextrust}{MPC-based \extrust}
\def\change#1{\ifchange\textcolor{blue}{#1}\else#1\fi}

\newcommand{\tikzhuman}[3]{
	\node[circle,fill=#3,minimum size=5mm,#2] (head#1) {};
	\node[rounded corners=2pt,minimum height=1.3cm,minimum width=0.4cm,fill=#3,below = 1pt of head#1] (body#1) {};
	\draw[line width=1mm,round cap-round cap, color=#3] ([shift={(2pt,-1pt)}]body#1.north east) --++(-90:6mm);
	\draw[line width=1mm,round cap-round cap, color=#3] ([shift={(-2pt,-1pt)}]body#1.north west)--++(-90:6mm);
	\draw[thick,white,-round cap] (body#1.south) --++(90:5.5mm);
}

\newcommand{\reqyes}{\textcolor{darkgreen}{\checkmark}}
\newcommand{\reqno}{\textcolor{darkred}{\xmark}}
\newcommand{\reqmaybe}{\textcolor{darkyellow}{(\checkmark)}}

\begin{document}



\title{\extrust: Reducing Exploit Stockpiles with a Privacy-Preserving Depletion System for Inter-State Relationships\iffullversion$^*$\thanks{* Please cite the journal version published at IEEE Transactions on Technology and Society 2023~\cite{RKGSR23}.}\fi}
\ifanonymous
\author{\textit{Anonymous Author(s)}}
\else
\author{%
    \IEEEauthorblockN{Thomas Reinhold\IEEEauthorrefmark{1}, Philipp Kuehn\IEEEauthorrefmark{1}, Daniel G\"{u}nther\IEEEauthorrefmark{2}, Thomas Schneider\IEEEauthorrefmark{2} and Christian Reuter\IEEEauthorrefmark{1}}
    
    \IEEEauthorblockA{\IEEEauthorrefmark{1}Science and Technology for Peace and Security (PEASEC)\\
        Technical University of Darmstadt, Germany} \\
        
    \IEEEauthorblockA{\IEEEauthorrefmark{2}Cryptography and Privacy Engineering Group (ENCRYPTO) \\
        Technical University of Darmstadt, Germany}
}
\fi

\begin{acronym}
\acro{nvd}[NVD]{national vulnerability database}
\acro{cve}[CVE]{common vulnerabilities and exposures}
\acro{cwe}[CWE]{common weakness enumeration}
\acro{cpe}[CPE]{common platform enumeration}
\acro{mpc}[MPC]{multi-party computation}
\acro{cvss}[CVSS]{common vulnerability scoring system}
\end{acronym}

\maketitle
\begin{abstract}
	Cyberspace is a fragile construct threatened by malicious cyber operations of different actors, with vulnerabilities in IT hardware and software forming the basis for such activities, thus also posing a threat to global IT security.
	Advancements in the field of artificial intelligence accelerate this development, either with artificial intelligence enabled cyber weapons, automated cyber defense measures, or artificial intelligence-based threat and vulnerability detection.
	Especially state actors, with their long-term strategic security interests, often stockpile such knowledge of vulnerabilities and exploits to enable their military or intelligence service cyberspace operations.
	While treaties and regulations to limit these developments and to enhance global IT security by disclosing vulnerabilities are currently being discussed on the international level, these efforts are hindered by state concerns about the disclosure of unique knowledge and about giving up tactical advantages. 
	This leads to a situation where multiple states are likely to stockpile at least some identical exploits, with technical measures to enable a depletion process for these stockpiles that preserve state secrecy interests and consider the special constraints of interacting states as well as the requirements within such environments being non-existent.
	This paper proposes such a privacy-preserving approach that allows multiple state parties to privately compare their stock of vulnerabilities and exploits to check for items that occur in multiple stockpiles without revealing them so that their disclosure can be considered.
	We call our system \extrust\ and show that it is scalable and can withstand several attack scenarios.
	Beyond the intergovernmental setting, \extrust\ can also be used for other zero-trust use cases, such as bug-bounty programs.
\end{abstract}


\section{Introduction}
\label{cha:intro}
The threat of malicious cyber activities is omnipresent and state actors are becoming an increasingly important part of this development~\cite{reuter_information_2019-1, Giles2019}, either due to the progressing militarization of cyberspace~\cite{lewis2013cyber, Koch2019} or due to cyber espionage operations~\cite{Buchan2018, Georgieva2019}.
At the same time, advancements in the field of artificial intelligence (AI) are being used to automate cyber defence measures~\cite{dhir2021prospective}, to develop AI enabled cyber weapons~\cite{Reinhold2022}, or to detect and predict software threats and vulnerabilities~\cite{Russell2018, Amarasinghe2019}.
In particular, knowledge of vulnerabilities is an integral part in most of these cyber operations \change{to} breach foreign IT-protection measures, and intelligence services and military forces stockpile such critical information without disclosing it for rectification~\cite{ablon_zero_2017, Rovner2020}. 
However, any serious and capable exploit withheld by a state for its own purposes becomes a potential threat for everyone, including the state itself, its economy, and civil society~\cite{NATOCCDCOE2020} as the exploit \emph{EternalBlue} exemplified in 2017~\cite{Reinhold2018, Cimpanu2019}.

One way out of this dilemma is a so-called vulnerability equity process (VEP)~\cite{Randal2019}, an institutionalized measure to regularly assess the criticality of stockpiled exploits and vulnerabilities to (re)consider their disclosure that could take place under the leadership of extra-national entities, such as the United Nations (UN)~\cite{Schulze2019}.
A major obstacle for such an approach is the reluctance of participating parties to disclose sensitive information about their own capabilities, which is generally seen as giving up tactical advantages, effectively resulting in an international arms race for offensive cyber capabilities~\cite{Harknett2020}. 
Historically, such situations have been countered by efforts to reach mutual agreements between states on arms control and reduction measures, \ie, treaty-based agreements to limit the risks of proliferation of weapon-enabled technology, to prevent its use with potentially disastrous consequences, or to reduce the risks of conflict arising by mistake or technological failures~\cite{reinhold_arms_2019}.

\change{With regard to the depicted development in cyberspace, early political approaches to mitigate these threats have been proposed by the UN~\cite{un-gge_report_2015}, the OSCE~\cite{osce_osce_2016}, and other organizations. But although first important steps towards an effective cyber arms control, like the exchange of threat information~\cite{sauerwein_threat_2017,Kuehn2020}, have been established, they are not suitable for limiting or reverting the aforementioned international cyber arms race of vulnerability stockpiling.}
So far, no proposal focuses on this specific challenge and the particular constraints of state actors, with their requirements of confidentiality, their potential mutual mistrust, and individual security concerns~\cite{Reinhold2019}.
\\

In this paper, we propose a technical solution called \extrust\ based on a multi-party computation approach that allows multiple actors to compare vulnerability stockpiles for matching entries while preserving their confidentiality. 
This includes an approach for the unique machine-readable identification of exploits that allows them to be checked for matches.
Our solution is designed for a zero-trust environment and does not rely on any preconditions of trust in advance or assumptions of good nature. 
This contributes to the development of measures for an international agreement to deplete vulnerabilities while circumventing the problems and impediments of intergovernmental cooperation. 

Beside this contribution, this paper further aims to provide an example of how politics is -- sometimes -- in need of technical solutions, in this case even for challenges regarding international security. 
As computer scientists and engineers are the experts on the domain of cyberspace, shaping it by developing software or even defining its constraints and rules themselves, we would like to encourage taking the responsibility that this entails seriously and support the peaceful development of this globally shared domain.



The paper is structured as follows:
Subsequent to this introduction, \cref{cha:relatedwork}~presents related work and elaborates the research gap.
\cref{cha:concept}~analyzes the requirements of \extrust\ both on a conceptual and IT security level.
\cref{cha:vulnid}~discusses how vulnerabilities can be uniquely described in a machine-readable form \change{that allows their comparison}.
\cref{cha:blockchain_approach}~presents {a Blockchain-based prototype approach that exemplifies the intended system and its requirements and discusses the challenges for a applicable \extrust\ implementation}.
\cref{cha:mpc_approach}~presents our contribution of a privacy-preserving exploit depletion system for zero-trust relationships using multi-party computation.
\cref{cha:discussion_and_application}~discusses the approach and evaluates it against the requirements. It also presents different application scenarios beyond state actors.
\cref{cha:conclusion}~concludes this paper and provides directions for future work.
In order to maintain readability, the technical details can be found in the Annex in~\cref{cha:annex}.
\section{Related Work}
\label{cha:relatedwork}
Since our paper covers and combines different computer science topics, this section summarizes the existing work on malware identification~(\cref{sec:terminology}), vulnerability mitigation~(\cref{sec:externaldepletion}), and promising cryptographic protocols~(\cref{sec:protocols}). 
Based on these descriptions, the research gap is described~(\cref{sec:researchgap}), which is closed by our approach.

\subsection{Vulnerability Terminology and Malware Identification Methods}
\label{sec:terminology}
An important prerequisite for comparing exploits -- as the core of a depletion system -- is the ability to create deterministic vulnerability descriptions.
Early attempts were based on the creation of so-called malware \emph{signatures}~\cite{cohen1987computer}, which function like a \emph{fingerprint}. Current malware detection approaches use a different approach that is either based on the entire binary code of the malware, \ie, the exploit and payload~\cite{Kirat:2015:MAE:2810103.2813642}, the compromised storage to create signatures~\cite{Petrik:2018:TAO:3243734.3278527} or even artificial intelligence measures to automatically generate descriptions via a Common Vulnerability Scoring System (CVSS) prediction ~\cite{kuehn_common_2023}.
Beside the actual detection of malware, other research area focuses on the description and identification of exploited vulnerabilities.
The popular \ac{nvd} provides a semi-structured database of known vulnerabilities~\cite{mitre_common_2005}, however, \etal{Dong}~\cite{dong2019towards} showed that the \ac{nvd} entries are inconsistent compared to other vulnerability databases.
Compared to the \ac{cve}, the \ac{nvd} entries differ in their announced project names or versions.
Alternative approaches were introduced by \etal{Sadique}~\cite{sadique_automated_2018} with the \emph{Structured Threat Information eXpression~(STIX)} and the \emph{Vocabulary for Event Recording and Incident Sharing Framework} (VERIS)~\cite{veris_vocabulary_2019} that can be used to describe, share, and publish threat information.
Both definitions, STIX and VERIS, offer a syntax for different types of threats, including malware, exploits, and vulnerabilities.
Some entry fields in \ac{nvd}, STIX, and VERIS may contain unstructured information that undermines unique descriptions.
\etal{Martin}~\cite{martin_cwe_2011} propose the \ac{cwe}, a dictionary of weakness classes that can be used to classify vulnerabilities, an approach we use in \cref{cha:vulnid} to identify vulnerabilities. 

\subsection{Vulnerability Mitigation \& External Depletion Measures}
\label{sec:externaldepletion}
Vulnerability research and mitigation methods have been a topic in IT security for several decades~\cite{one1996smashing, you_semfuzz_2017, carlini_rop_2014, shacham_geometry_2007}.
One measure are so-called \emph{bug-bounty programs}~\cite{Zhao:2014:ESW:2663887.2663906} like e.g. \emph{HackerOne}~\cite{perlroth_hackerone_2015}, which aim to attract IT security practitioners to penetrate advertised systems and services and report loopholes in software or services. 
Other programs are run by Mozilla, Facebook, and Microsoft~\cite{mozilla_security_2017,facebook_facebook_2018,zimmerman_microsoft_2017} or \emph{Project Zero}~\cite{evans_announcing_2014} by Google, which focuses on the search for zero-day vulnerabilities.
These programs, which we further refer to as \emph{external depletion measures,} aim to identify vulnerabilities in popular IT products \change{to} disclose them to the producers and get them fixed as a depletion measure.

\change{In contrast, \emph{internal depletion measures} focus on an actor's secret exploit stockpile of already known, but not yet disclosed vulnerability information. 
Such measures have not yet been proposed before for the given application context of interstate cooperation and international security.}

Practical approaches at this international, intergovernmental level have so far been limited to transparency and confidence-building, rather than arms control and the non-proliferation or disarmament of malicious cyber tools.~\cite{reinhold_arms_2019}.

\subsection{Cryptographic Protocols}
\label{sec:protocols}
Our \extrust\ system is related to well-studied cryptographic protocols like multi-party computation~(cf.~\cref{sec:mpc}), private set intersection~(cf.~\cref{sec:crypto}), and trusted hardware~(cf.~\cref{sec:hardware}). These approaches are further elaborated in the following.

\subsubsection{Multi-Party Computation (MPC)}
\label{sec:mpc}
The first approaches to \ac{mpc} of functions represented as a Boolean circuit were proposed by Yao~\cite{Yao1986} for $N=2$~parties with constant round complexity, and by Goldreich, Micali, and Wigderson~(GMW)~\cite{Goldreich1987} for any number of parties~$N$ with round complexity linear in the depth of the Boolean circuit.
Beaver, Micali, and Rogaway~(BMR)~\cite{Beaver1990} extended Yao's protocol to the multi-party case while maintaining the linear round complexity.
Based on this initial work, many research projects followed, showing the practical feasibility of \ac{mpc} for many privacy-preserving applications, such as auctions~\cite{Bogetoft2009}, set intersection~\cite{PinkasSZ18}, and machine learning~\cite{Mirhoseini16}.
\etal{Kamara} presented an outsourcing technique~\cite{Kamara2011}, which allows~$N$ parties to outsource the \ac{mpc} protocol to $n \ll N$~parties. 

\subsubsection{Private Set Intersection (PSI)}
\label{sec:crypto}
Private Set Intersection (PSI) has been proposed to identify malware~(cf.~\cref{sec:terminology}) in a single client and server environment~\cite{Kiss2017}.
A recent survey and performance comparison of different PSI protocols by \etal{Pinkas}~\cite{PinkasSZ18} demonstrates that the approach proposed by Pinkas, Rosulek, Trieu and Yanai ~\cite{PaXoS} is currently the fastest PSI protocol which can handle malicious security.
In our proposed application context, we have multiple parties, hence we are mainly interested in multi-party PSI.
Multi-party PSI protocols with passive security are applied by \etal{Kolesnikov}~\cite{Kolesnikov2017} and \etal{Inbar}~\cite{10.1007/978-3-319-98113-0_13}.
A scalable, maliciously-secure multi-party PSI protocol is presented by Hazay and Venkitasubramaniam~\cite{Hazay2017}.
\etal{Huang}~\cite{Huang2012} use a general \ac{mpc} framework to privately compute the set intersection between two parties.

\subsubsection{Trusted Execution Environment (TEE)}
\label{sec:hardware}
Another promising approach for a privacy-preserving exploit depletion system is to securely isolate the execution into a \emph{trusted execution environment} (TEE)~\cite{anati2013innovative}, that allows untrusted data to be computed in a secure environment that is isolated from all other executions running on the same machine, where it is protected against manipulation and disclosure.
TEEs are omnipresent in all Intel processors from the 6th generation upwards as~\emph{Intel Software Guard Extension}~(SGX).
Although many works use Intel SGX for efficient secure multi-party computation~\cite{Koeberl2015, Gupta2016, Kucuk2016, Bahmani2017, Felsen2019}, TEEs are not suitable for applications when states are involved, since this would require that state actors trust the hardware-producing countries not to manipulate the TEEs, \eg, by including backdoors.

\subsection{Research Gap}
\label{sec:researchgap}

Above all, practical measures are a mandatory aspect of potential arms control and disarmament treaties, as history and insights into former weaponized technologies have shown~\cite{goldblat2002arms}. Existing IT methods such as Multi-PSI~\cite{Hazay2017}~(cf.~\cref{sec:crypto}) and secure hardware~\cite{Felsen2019}~(cf.~\cref{sec:hardware}) have not been applied to exploit depletion, especially regarding the demands and particular constraints of an interstate zero-trust environment.
Such a protocol for pairwise PSI among~$N$ parties, as required for a privacy-preserving exploit depletion system, is currently not available. Thus, our approach \extrust\ proposes a Boolean circuit that implements the desired functionality via \ac{mpc}~(cf.~\cref{cha:mpc_approach}).

\section{Requirements Analysis}
\label{cha:concept}
In this section, the requirements of \extrust\ are analyzed as a system for reducing exploit stockpiles, resulting from the chosen context of interstate relations. This list is divided into conceptual requirements derived from the specific constraints of the context of arms control, as well as the IT security requirements in combination with the selection of the adversary model.

\subsection{Conceptual Requirements}
\label{sec:requirements}
As mentioned above, this paper focuses on cases in which two or more parties stockpile vulnerabilities and exploits. This reflects the character of arms control treaties, whose ``practical'' part of active mutual control or (limited) cooperation measures are always based on bi- or multilateral agreements~\cite{Reinhold2019b} between a small group of states.
Based on a rational choice consideration~\cite{Zangl1994}, our approach builds upon the following two premises, that we consider to be reflected by states that stockpile vulnerabilities~\cite{Kraus2019}, as they resemble the considerations behind a vulnerability equity process~\cite{Randal2019}. \change{Firstly, we consider states to be aware, that withholding a vulnerability poses a potential threat to their own IT systems. Secondly, we consider that a vulnerability which is known to more than one state is more likely to be considered a candidate for disclosure, because its intended effect is probably ineffective or at least uncertain and because disclosing the vulnerability results in publicly available security patches that support the state's own IT security and also renders the vulnerability worthless for everyone else.}


On the other hand, all vulnerabilities are high-value assets for the stockpiling party. Given the context of state interaction, each party will try to avoid revealing any information that can lead to the loss of tactical advantages, while trying to extend these advantages by gaining information about the other parties. In addition, arms control measures are established in times of political tensions \change{to} avoid the outbreak of armed conflict. Based on these assumptions, we consider that \extrust\ has to operate in a zero-trust environment in which parties have to be incentivized to cooperate, while at the same time assuming that other parties are either extremely reluctant to disclose information, attempt to gain information for their own interest, or are \change{otherwise dishonest regarding their cooperation and activities.}

With these considerations in mind, \extrust\ aims to require as little cooperation as possible due to this zero-trust environment. This means that each party discloses only the absolutely necessary amount of information, thereby retaining all specific information about capacities and capabilities. 
Additionally, each party should be able to perform its own check for intersections at any time without relying on further cooperation, dedicated data exchange, or any form of super-ordinate institution. 
Furthermore, information already provided should not be allowed to be altered, deleted, or corrupted.

In light of this context, the necessary measure needs to fulfil the following conceptual requirements (RC):
\begin{enumerate}[label=RC\arabic*, leftmargin=2.4em]
	\item The measure has to enable parties to add information about vulnerabilities and exploits.
	\item Intersection checks have to be able to be performed by either party at any time without having to obtain the consent of the other parties involved. A match is considered as such if at least two different participating parties have submitted identical information about vulnerabilities or exploits.
	\item The system has to send feedback when it detects an intersection match. 
	\item Although real-time computability is not strictly necessary for processes that are usually politically slow, such as arms control measures, the system needs to be scalable with respect to the number of parties so that parties can join or leave at any time. While previous arms control treaties are usually established in a small circle of state actors that participate in mutual control measures, indicating there could be up to~$N=5$ participating parties in a real-world arms control scenario, this should not be the upper bound of our system.
	\item The system should be operated decentralized and not require a specific neutral authority to operate or maintain the system.
\end{enumerate}

\subsection{Adversary Model}
\label{sec:adversary}
The two most common adversary models are semi-honest (passive) and malicious (active) adversaries~\cite{EKR18}.
While semi-honest adversaries follow the underlying rules and procedures (in technical terms the so-called~\emph{protocol}) and try to extract as much information as possible from the transcript, malicious adversaries may arbitrarily deviate from the agreed rules.
Given the zero-trust environment in the context of \extrust,\ we consider an active or malicious attacker as adversary model.
Although technical security measures that protect against semi-honest adversaries are more efficient than those against malicious adversaries, we must consider state actors that might maliciously manipulate arms control computations and outcomes.
Additionally, we assume a dishonest majority, \ie, up to~$N-1$ parties may be malicious.
The motivational scenario of \extrust\ is a highly security critical one in which top secret information may be exchanged.
Hence, it should withstand several passive attacks, like eavesdropping, and also be shielded against active attacks, such as flooding or brute-force attacks.
We have therefore chosen the model of the stronger adversary in contrast to the passive, semi-honest adversary.
This decision also covers the application context of the zero-trust relationship between the actors involved.

\subsection{Technical and Security Requirements}
\label{sec:security_requirements}
In addition to the conceptual requirements, the approach must meet additional security expectations to provide an applicable and secure measure of exploit depletion in a zero-trust environment. 
The requirements reflect the need for confidentiality and are important to motivate stakeholders to participate.
These technical and security requirements (RS) are:
\begin{enumerate}[label=RS\arabic*, leftmargin=2.4em]
	\item The system must ensure the confidentiality of vulnerability or exploit information against any party. 
	\item Submitted data should not be able to be withdrawn, modified, or corrupted by any party.
	\item The system needs to prevent false positive intersection results.
\end{enumerate}

In the following, after discussing the identification of vulnerabilities as a necessary prerequisite of our system, we present a prototype solution for \extrust\ that addresses these requirements and illustrates its inherent challenges.
Afterwards, we present our contribution of a \mpcextrust.
\section{Identifier of Vulnerabilities}
\label{cha:vulnid}
In this section, we propose a unique, machine-readable identification method for vulnerabilities to be able to match them. The mathematical description of the required properties and the associated challenges can be found in the Annex and are referenced here.

\subsection{Machine-Readable Vulnerability Identifier}
At its core, \extrust\ privately matches vulnerabilities or exploits of different parties.
This requires using a vulnerability description method that results in the same machine-readable descriptions for the same vulnerability\footnote{See \iffullversion \cref{sec:vulnerabilityIdentifier}. \fi}.
An established approach to describe and thus identify vulnerabilities is provided by vulnerability databases like the \ac{nvd}.
The \ac{nvd}'s entries, for example, contain information used for identification. Their semi-structured format, however, makes it practically impossible for individuals to independently create the same identifier for a vulnerability.
Therefore, we use the approach of \etal{Kuehn}~\cite{Kuehn_Bayer_Wendelborn_Reuter_2021} to achieve uniqueness, \ie, we adjust the \ac{nvd}'s entry information by removing any free-form pairs and pairs that provide no information about the vulnerability itself~(\eg,~the~CVE-ID), align the structured information with the vulnerability descriptions, and add information about the vulnerable function, extracted from the vulnerability description.

\begin{figure}[t]
\begin{lstlisting}[language=json,caption=Vulnerability Identifier for CVE-2020-28877,label=lst:vulnid]
{
  "cpe": "cpe:2.3:o:tp-link:wdr7400_firmware:-:*:*:*:*:*:*:*",
  "cwe": 120,
  "fun": "copy_msg_element"
}
\end{lstlisting}
\end{figure}

The remaining fields are \ac{cwe} and \ac{cpe} with the addition of the vulnerable function, which are structured and algorithmically comparable.
The \ac{cwe}~\cite{mitre_cwe_2019} defines hierarchical layers of vulnerability weakness classes, while the \ac{cpe}~\cite{cpe} provides a machine-readable way to describe platforms.
If a vulnerability affects multiple platforms, we use separate vulnerabilities for each affected platform.
The resulting vulnerability identifier is depicted in~\cref{lst:vulnid}~(for~CVE-2020-28877).

\subsection{Analysis}
\label{subsec:limitationsidentifier}
Using a simple object notation for the vulnerability identifier offers flexibility and extensibility, and by adding \ac{cpe} and \ac{cwe} as well as the vulnerable function as core elements, identifiers can be specific enough to create matching values when different actors describe and submit the same vulnerability or exploit.
This is essential to identify matching vulnerabilities.

The main limitation of the vulnerability identifier's definition is based on a trade-off between the properties \emph{accuracy} and \emph{ambiguity}.
Currently, it is still possible to describe two different vulnerabilities with the same identifier, or to use two different identifiers for the same vulnerability\footnote{See \iffullversion \cref{sec:ambiguousVulnerabilityIdentifier} \fi}.
This leads to false positives ~(two different vulnerabilities are mapped to one identifier) or false negatives ~(the same vulnerability is mapped to two different identifiers), respectively, depending on the level of detail implemented into the identifier.
However, there are possibilities to adjust the identifier definition accordingly.
Increasing the amount of information captured by the identifier makes the identifier more specific but introduces more ambiguity, \ie, false negatives.
Parameters to be added are the \ac{cvss} parameter information~(\eg, impact information) or the vulnerable path~(\ie, the filename in which the vulnerability resides)~\cite{Kuehn_Bayer_Wendelborn_Reuter_2021}.
Another way to adjust the identifier is the \ac{cwe}'s hierarchy depth.
\ac{cwe} classes are hierarchically ordered and thus offer generalization or specification.
Including relations of the used \ac{cwe} class increases the specificity of the identifier and could help to circumvent cases where identifiers use different \ac{cwe} subclasses of the same top level class.
\
At this point, we want to stress that in the presented scenario~(cyber arms control) false positives must be avoided, while false negatives are tolerable.
If false positives are a common problem in such a system, it would drastically lose acceptance among states that are still interested in stockpiling vulnerabilities.

The size of the proposed identifier space is restricted by the number of \ac{cwe} classes, the size of the \ac{cpe} directory, and the possible function names, which serve as secret information.
Individually, these spaces can be approximated in their size.
For the space of possible function names~$\mathit{FN}$, we assume a clean coding style, \ie, function names are descriptive and use at most three English words with any kind of connector (\eg, camel case or underscores), which results in $\approx 2^{81}$ identifiers\footnote{See \iffullversion Section~\ref{sec:vulnerabilityIdentifierSpace}. \fi}.

As argued, the presented approach is sufficient to describe vulnerabilities uniquely. It serves our needs with a trade-off in detail that avoids both different vulnerabilities being described by the same identifier as well as the same vulnerability being described with different identifiers. 
Based on the current limitation of the identifier space, brute-force attacks remain a problem and efforts should be made to increase the identifier space.  
As an alternative to the proposed definition of identifiers, our system \extrust\ can work with any other scheme that is concise, structured, and unambiguous.

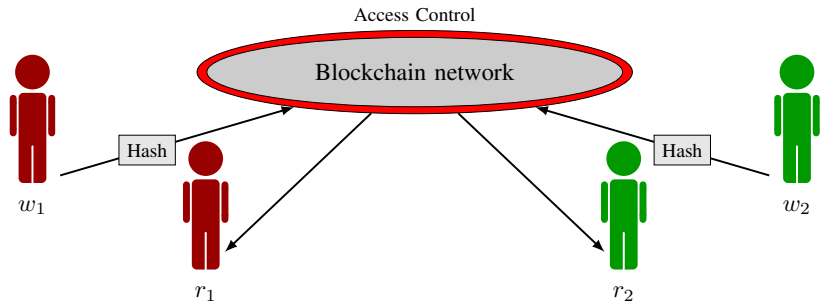
\begin{figure*}[ht]
	\centering
	\resizebox{0.6\textwidth}{!}{%
		\begin{tikzpicture}[%
				node distance=1cm]
			\node (network) [draw, ellipse, fill=black!20, text centered, text width=4cm, minimum height=1cm] {Blockchain network};

			\tikzhuman{leftreader}{below left=of network}{red!60!black};
			\node (dlr) [below of=bodyleftreader] {$r_1$};
			\tikzhuman{rightreader}{below right=of network}{green!60!black};
			\node (drr) [below of=bodyrightreader] {$r_2$};

			\node (stepleft) [left=of network] {};
			\node (stepright) [right=of network] {};

			\tikzhuman{leftsubmitter}{left=of stepleft}{red!60!black};
			\node (dls) [below of=bodyleftsubmitter] {$w_1$};
			\tikzhuman{rightsubmitter}{right=of stepright}{green!60!black};
			\node (drs) [below of=bodyrightsubmitter] {$w_2$};

			\node (hashleft) [draw,fill=black!10,font=\footnotesize] at ($(bodyleftsubmitter.south)!0.3!(network)$) {Hash};
			\node (hashright) [draw,fill=black!10,font=\footnotesize] at ($(bodyrightsubmitter.south)!0.3!(network)$) {Hash};

			\begin{pgfonlayer}{background}
			\node (accesscontrol) [draw, ellipse, fill=red, text centered, text width=4.2cm, minimum height=1.2cm] {};
			\node (descriptionaccesscontrol) [font=\footnotesize, above=.1cm of network] {Access Control};

			\node (leftreaderstep) at ($(bodyleftreader.south)!0.05!(accesscontrol)$) {};
			\node (rightreaderstep) at ($(bodyrightreader.south)!0.05!(accesscontrol)$) {};
			\node (leftsubmitterstep) at ($(bodyleftsubmitter.south)!0.05!(accesscontrol)$) {};
			\node (rightsubmitterstep) at ($(bodyrightsubmitter.south)!0.05!(accesscontrol)$) {};

			\draw [->, >=latex, thick] (accesscontrol) -- (leftreaderstep);
			\draw [->, >=latex, thick] (leftsubmitterstep) -- (accesscontrol);
			\draw [->, >=latex, thick] (accesscontrol) -- (rightreaderstep);
			\draw [->, >=latex, thick] (rightsubmitterstep) -- (accesscontrol);
			\end{pgfonlayer}
		\end{tikzpicture}
	}

	\caption{System architecture for \bcextrust. $r_i$ and $w_i$ denote readers and writers of actor~$i$.}
	\label{fig:system}
\end{figure*}

\section{\extrust\ using Blockchain}
\label{cha:blockchain_approach}
\change{To} illustrate the challenges involved in implementing a privacy-preserving exploit depletion system, we have chosen a simple, straightforward prototype based on a Blockchain implementation, referred to hereafter as \emph{\bcextrust}. 
Although this approach entails security flaws from a theoretical perspective, we want to use this prototype to illustrate, test, and analyze possible solutions regarding the requirements and the proposed depletion process, as an introduction for our multi-party computation-based approach presented in \cref{cha:mpc_approach}.
This section presents the architecture and proof-of-concept implementation of this prototype and concludes with a discussion of the requirements met as well as the identified constraints.    

\subsection{System Architecture and Procedure}
\label{sec:blockchain_architecture}
In terms of conceptual requirements, \bcextrust\ should run in a \emph{distributed setting with no central trusted authority}, with a \emph{complete, secured, and tamper-resistant history} of all submitted information and should allow \emph{asynchronous intersection checks} that can be performed by each participating party independently. 

We have developed a prototype based on a private Blockchain technology~\cite{Gupta2017,zheng_blockchain_2018} that provides all of these features. A private Blockchain is a distributed chain of \emph{blocks} containing \emph{transactions}, where each block references its previous block via hard-to-calculate mathematical challenges and cryptographic hashes to reference the block. 
This provides a tamper-proof history of all submissions, as any modification would invalidate the adjacent entries.
The data storage part of a Blockchain, the so-called ledger, is replicated to all participants and automatically synced between them. In private networks, access to it is walled by an access control manager~(ACM). 
The interfaces for interaction with the ledger are called \emph{smart contracts}.
With regard to the system architecture, the ledger provides the storage space, the smart contract is responsible for the submission and comparison mechanism, and the ACM controls the access as well as the different layers of interaction permissions via roles and associated authorizations.
To maintain the confidentiality of the submitted vulnerability identifiers, we secured the information using cryptographic hash functions~\cite{katz_introduction_2014}.

The overall procedure begins with the setup of the Blockchain instance (\emph{nodes}) by each participating party and their interconnection to build an evenly distributed network.
\change{To} submit a vulnerability, the vulnerability identification method we propose in \cref{cha:vulnid} is used to create an identifier for the specific vulnerability, which is then cryptographically secured using a hash function and finally stored in \bcextrust.
Afterwards, any participating party can perform a transaction, which checks for intersections between all hashes stored in the ledger and logs the output on the ledger.
This way, a history of all actions performed is ensured, which is accessible to any involved party, including intersections.
Nevertheless, parties that do not know the plaintext vulnerability identifier cannot obtain any information other than the fact that an intersection occurred.

\subsection{Implementation}
\label{cha:blockchain_implementation}
To focus on developing a proof-of-concept implementation of \bcextrust, we decided to utilize a private Blockchain framework~\cite{davies_private_2018} as it provides all relevant tools for the interaction of the actors with the system, the data structures for storing information, and all necessary data operations for reading, writing, and verifying information within the stored data. 
We have selected the \emph{Hyperledger Fabric}~\cite{voell2016hyperledger} Blockchain framework because it is open source, actively maintained and well documented, and provides the rapid prototyping environment \emph{Hyperledger Composer}~\cite{hyperledger_hyperledger_2017}  with a boilerplate implementation for each part of the Blockchain network.

With regard to the permissions of participants using \bcextrust, we envision two roles: Readers, who can read the entire ledger and perform the transaction that checks for matching items; and writers, who can only submit items (see \cref{fig:system}). This restriction is only necessary due to the use of our chosen framework\footnote{As framework, we chose Hyperledger Composer.}, otherwise, a party's submission may be intercepted and copied by other parties.
The theoretical concept does not require this separation, because no party would be able to access other parties' information.

The items which are submitted and stored into the ledger are the vulnerability identifiers, as described in (\cref{cha:vulnid}). 
As the plain vulnerability identifier must never be inserted into the Blockchain network to prevent its exposure to all parties involved, it is obscured before being submitted.
We generate a cryptographic hash of a normalized JSON representation of the vulnerability identifier via \texttt{SHA3-512} (\cite{sha512}), following the NIST's policy on hash functions~\cite{nisthash}.
This provides a 256-bit security level. 

To interact with \bcextrust, the prototype system provides two transactions: The simple submission of hashed vulnerability identifiers and the transaction that checks the stored hashes within the ledger and triggers an event along with references to matching items, \texttt{checkIntersections}\footnote{See \iffullversion \cref{sec:checkIntersections}. \fi}.  

We want to stress that this prototype implementation does not yet take performance into account, as this is no core requirement of \extrust and its proposed arms control application.

\subsection{Discussion of \bcextrust}
\label{cha:blockchain_limitations}

As indicated earlier, the development of IT measures is a novel approach in the field of technical tools for cyber arms control that has to balance conflicting objectives to a certain extent. 
For arms control, the aspect of minimum requirements for cooperation between the parties is essential, as it establishes the lowest possible barrier for participation. 
This is crucial for situations such as the intended one, in which trust cannot be assumed as a given motivation for cooperation. 
In previous treaties, this often meant a certain degree of pragmatism regarding the acceptance of ``gray areas'' and the possibility of non-compliance. 
The opposite objective is the requirement of technically secure solutions, as this too provides important incentives for participation.
This in turn is likely to result in protocol specifications creating operational conditions that potential participants are not prepared to accept.

Considering the requirements, the Blockchain approach provides a manipulation-proof and distributed storage of all submitted information.
Calculations are distributed and performed independently, thereby mitigating the need for a trusted third party to maintain the shared information, as well as any other form of cooperation beyond the actual submission.
The system can include additional parties without adjustments or significant impact on the performance of the system, beyond the network capacity necessary to synchronize the stored information~\cite{Nadeem2019}. 
In addition, the processing of submissions is not time-critical, which is considered a bottleneck for massive, high-traffic Blockchain applications~\cite{Chohan2019, Croman2016}.
By securing vulnerability identifiers, the confidentiality of the information is -- at least theoretically -- maintained both in submission and in intersection detection.

On the other hand, the Blockchain-based prototype has serious IT security issues, both for active attackers (like non-participating state parties that try to break the system \change{to} gain advantages and reveal secret information) and fraudulent, semi-honest state participants that try to gain information which goes beyond the agreed exchange. 
Notably, the information contained in the distribution of the ledger is vulnerable to brute-force attacks by testing hashes, as foreign countries could generate possible vulnerability descriptions and test them against their local ledger.
The PSI literature has demonstrated that private elements cannot be hidden by simple hashing~\cite{Demir2018, Kales2019}.
The probability of creating an existing hash is based on the size of the identifier space and influenced by the number of its properties and values. 
As the identifier space of \bcextrust\ is very small (28 bit, cf.~\cref{subsec:limitationsidentifier}) brute-force attacks are very efficient and can be successfully exploited.
In addition, the brute-force attack is completely local since states have a local copy of the whole ledger. Consequently, states would not even notice if a brute-force attack was exploited to find all ledger vulnerabilities.
The brute-force attack can be slowed down (but not prevented) by using a difficult to parallelize hash function such as \emph{Argon2}~\cite{biryukov2016argon2}.
Extending the identifier space for the vulnerability by more complex identifier descriptions is not an option either, as this increases the probability of describing the same vulnerability differently. 

The Blockchain also faces other attack scenarios, such as the so-called 51\% attack, which allows attackers to manipulate the ledger~\cite{8563187}.
Attackers could also use more subtle ways to create intersections to test foreign submissions by creating \emph{fictional vulnerabilities} for rare software systems based on clever and informed guesses. 
This could also be used for targeted vulnerability suppression if a participating party creates and submits specific vulnerabilities, intentionally wrongfully signalling its possession \change{to} force the vulnerability to be disclosed.
In addition, a dishonest state party could clone and resubmit hashes under its own flag, which would also cause \bcextrust\ to false signal to the original submitter that this particular vulnerability can be eliminated.
However, such a cheat gives the attacker only a slight advantage, as they do not know what the cloned vulnerability information contains, and are likely to attract attention if performed regularly.
A final IT security issue concerns passive adversaries that gain access to the ledger, as well as the complete disclosure of the ledger to non-involved third parties.  
Besides the brute-force attack, the attacker will be able to learn which hashes belong to which party via timing correlations, detecting the amount of different participating actors as well as the amount of submitted hashed items stored by each actor.

The \bcextrust\ prototype has shown that it provides the conceptual requirements that arise from the arms control context. 
Regarding the attack scenarios described, it is important to emphasize that for this application, any attempt to attack or misuse the system is contrary to the principles of the confidence-building aspect of such a mutual measure and its political signal of de-escalation. 
It is further expected that all parties comply with the defined rules to at least achieve a positive outcome for their own national security. 
Nevertheless, this expectation needs to rest upon a secure protocol that inherently prevents fraud and guarantees the promised confidentiality. 

The following section presents the approach of \mpcextrust, an arms control measure that provides this level of security.
\begin{figure*}[htbp!]
	\centering
	\resizebox{0.6\textwidth}{!}{%
		\begin{tikzpicture}[%
				node distance=1cm]
			\node (network) [ellipse, text centered, text width=4cm, minimum height=1.5cm] {};

			\tikzhuman{leftreader}{below left=of network}{red!60!black};
			\node (dlr) [below of=bodyleftreader] {$p_2$};
			\tikzhuman{rightreader}{below right=of network}{green!60!black};
			\node (drr) [below of=bodyrightreader] {$p_3$};

			\node (stepleft) [left=of network] {};
			\node (stepright) [right=of network] {};

			\tikzhuman{leftsubmitter}{left=of stepleft}{blue!60!black};
			\node (dls) [below of=bodyleftsubmitter] {$p_1$};
			\tikzhuman{rightsubmitter}{right=of stepright}{black};
			\node (drs) [below of=bodyrightsubmitter] {$p_4$};


			\begin{pgfonlayer}{background}
			
			\node (readerleft) at ($(bodyleftreader)!0.1!(bodyleftsubmitter)$) {};
			\node (readerright) at ($(bodyleftreader)!0.05!(bodyrightsubmitter)$) {};
			\node (reader2left) at ($(bodyrightreader)!0.05!(bodyleftsubmitter)$) {};
			\node (submitterleft) at ($(headleftsubmitter)!0.03!(headrightsubmitter)$){};
            \node (submitterright) at ($(headrightsubmitter)!0.03!(headleftsubmitter)$){};
            \node (reader2right) at ($(bodyrightreader)!0.1!(bodyrightsubmitter)$) {};

			\draw [<->, >=latex, thick] (readerleft) -- (submitterleft);
			\draw [<->, >=latex, thick] (readerright) -- (submitterright);
			\draw [<->, >=latex, thick] (readerright) -- (reader2left);
			
			\draw [<->, >=latex, thick] (submitterleft) -- (submitterright);
			\draw [<->, >=latex, thick] (submitterleft) -- (reader2left);
			
			\draw [<->, >=latex, thick] (submitterright) -- (reader2right);

			\end{pgfonlayer}
		\end{tikzpicture}
	}

	\caption{MPC setting with four participating parties~$p_1, \dots, p_4$.}
	\label{fig:mpcsystem}
\end{figure*}
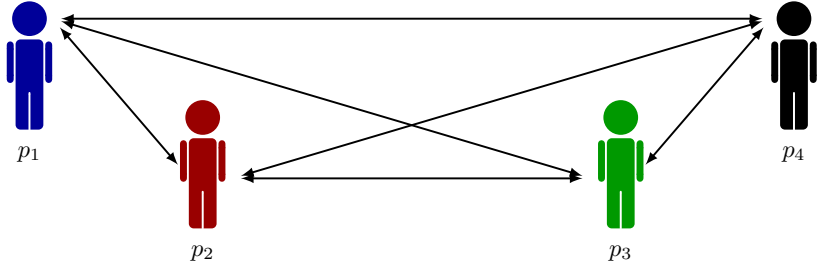

\section{\extrust\ using Multi-Party Computation}
\label{cha:mpc_approach}
This section presents our approach for an \mpcextrust to develop an exploit depletion system under the conditions of an untrusted environment that fulfils the discussed conceptual and security requirements  (\cref{cha:concept}), while avoiding the security problems that our prototype revealed (\cref{cha:blockchain_approach}). This approach is based on an interactive \emph{Multi-Party Computation} (\ac{mpc}) protocol as grounds for our \mpcextrust\ architecture. In the following, we present the concept and design of \mpcextrust\, technical details can be found in the Annex in section "PSI-variant Boolean circuit for multiple parties". \\
Secure \ac{mpc}~\cite{Archer18, Evans18} enables~$N$~parties to securely compute a commonly agreed public function~$f$ on their respective secret inputs~$x_1, \dots, x_N$ without revealing anything other than the result of the calculated function~$f(x_1, \dots, x_N)$.
\ac{mpc} guarantees that each of the~$N$ parties will not learn any information~(\eg,~input from the other parties or intermediate results of the computation) other than what a party would learn in the ideal world with a trusted third party.
In the ideal world, all parties send their inputs~$x_1, \dots, x_N$ secretly to a trusted third party, which then locally computes the function $f(x_1, \dots, x_N)$ and broadcasts the result to the~$N$~parties.
In the proposed context of arms control, even if such a trusted third party existed (\eg,~in the UN framework), it would probably not be accepted by all state actors or, at the very least, would raise the barrier to participation in the proposed measure (see \cref{sec:requirements}).

In \ac{mpc}, the function~$f$ that shall be computed is represented as a \emph{Boolean circuit}.
A Boolean circuit is a logical function whose operations are so-called \emph{Boolean gates}. A Boolean gate takes a set of Boolean inputs (\ie, either 0 or 1) and computes one Boolean output. We represent our \mpcextrust{} functionality as a Boolean circuit as efficient cryptographic protocols exist that can securely evaluate Boolean circuits.
A Boolean circuit consists of inputs, outputs, and Boolean gates that have two inputs and one output in the Boolean set $\{0, 1\}$.
The input of a gate can either be one of the inputs of the Boolean circuit or an output of a previous gate.
In \ac{mpc}, Boolean circuits usually only consist of \texttt{AND} and \texttt{XOR} gates, as any functionality can be realized using these two gate types.
A two-input \texttt{AND} gate outputs `1' if both of its inputs are set to `1', while a two-input \texttt{XOR} gate outputs `1' if exactly one (but not both) of its inputs is set to `1'.
For the actual algorithm that processes the submitted information and checks for collisions -- the so-called \emph{protocol} -- we use the BMR protocol~\cite{Beaver1990} and refer to Braun et al.~\cite{MOTION} for a detailed protocol description. We further use a well-established outsourcing technique~\cite{Kamara2011} to distribute the information processing for a group of~$N$ parties to $n \ll N$~parties. This setting for our \ac{mpc} approach is shown in~\cref{fig:mpcsystem}.
In summary, a subset of $n$ from $N$ (state) parties interactively run an MPC protocol on a Boolean circuit, which computes the functionality of \extrust.
This setting allows us to evaluate the functionality of \extrust{} in a privacy-preserving manner, while reducing the number of active parties that are fully involved in the computation ensures the scalability of \mpcextrust.

The protocol requires that the parties have sorted their inputs locally before they are fed into the \ac{mpc} protocol.
\change{To} verify this, we use the Boolean circuit and open intermediate values so that the parties can abort the protocol execution if a malicious party has not sorted its inputs correctly.
When opening these values, the remaining intermediate values before and after the opening process must be protected to allow further secure computation with them.
Many efficient maliciously-secure \ac{mpc} protocols provide this property, known as \emph{reactive \ac{mpc}}, \eg,~\cite{Damgard10, Bendlin11, Nielsen12, Damgard12, Damgard13}.
Apart from checking correctly sorted sets, we use reactive \ac{mpc} to maintain the state of the secretly shared inputs after the end of a protocol run, so that submitted vulnerabilities do not need to be secretly shared again in the next iteration.

This \ac{mpc} approach allows us to develop a privacy-preserving exploit depletion system that fulfils the requirements of the arms control context (\cref{cha:concept}).

\subsection{System Architecture}
\label{sec:mpc_architecture}

Our complete \mpcextrust\ architecture works as follows (see Fig.\ref{fig:mpcsystem}):
$N$~parties try to find intersections of their own identified vulnerabilities between themselves and at least one other actor.
These~$N$~actors securely evaluate a Boolean circuit~(cf.~\cref{sec:mpc}), consisting of  \texttt{AND} and \texttt{XOR} gates\footnote{See \iffullversion \cref{fig:mpccircuit}. \fi}, that takes as input the known vulnerabilities and exploits of the actors, which are represented as hash values~(cf.~\cref{cha:vulnid}), compares them to find intersections, and finally outputs all intersections found to the respective parties.
This circuit, however, is not constant over the lifetime of \mpcextrust\ as it depends on the number of parties~$N$~(states can be added/removed) and inputs~$u$~(vulnerabilities can be added).
The participating parties perform an initial~\ac{mpc} protocol prior to the actual execution to determine the maximum number of vulnerabilities~among all parties, which then determines the number of inputs for the Boolean circuit that is evaluated by the MPC protocol.
Now that every party knows the number of inputs to the Boolean circuit, each party in the fixed subset $n$ of the $N$ parties locally compiles the Boolean circuit that is inserted into the~\ac{mpc} protocol, \ie, no further interaction is required by the parties to agree on the Boolean circuit.
Malicious-secure~\ac{mpc} protocols ensure that parties, who compiled a fake Boolean circuit that does not compute the agreed functionality, are identified by the other parties. \change{Regarding the already submitted vulnerabilities, a party cannot revoke or modify submitted information because they remain in the input list of the~$N$~actors.}
The parties can opt in and out by sending a notification message to the~$N$~servers.
Only the inputs of the participating parties that are logged in are taken into account for the computation.
\newline
\subsubsection{Complexity and optimization of the Boolean circuit}
For~$N$~parties, state-of-the-art \ac{mpc} protocols require sending and receiving~$\mathcal{O}(N)$ messages for each \texttt{AND} gate in the Boolean circuit~\cite{Beaver1990}, while \texttt{XOR} gates can be computed locally without any interaction between the parties \cite{Lindell2016}.
Consequently, we optimize the number of \texttt{AND}~gates in our Boolean circuit that is evaluated via \ac{mpc}.
In order to prevent the concrete set sizes of the individual parties from being leaked, we specify an upper limit~$u$ that determines how many inputs a party feeds into the circuit.
If a party has fewer than~$u$~inputs, it fills the missing inputs with random dummy values, which will not represent any vulnerability and thus will not occur in any intersection as the probability that two parties independently choose the same random dummy values is negligible.
On a high level, every party inputs two unique keys -- $k_0$ and $k_1$ -- for each of its vulnerabilities into the Boolean circuit. The Boolean circuit outputs $k_1$ if this vulnerability is part of an intersection or $k_0$ if only the respective party knows this vulnerability.
Although the resulting keys are leaked to all parties, only the party who input the keys learns any information about the intersecting identifiers of their vulnerability.
\newline
\subsubsection{Using Private Set Intersection to calculate collisions}
\change{To} calculate the intersection of different stockpiles, we use the Private Set Intersection (PSI) protocol.
PSI allows two parties to securely compute the intersection of their private sets without leaking any information about set elements that are not part of the intersection to the other participating party.

Multi-party PSI~\cite{Hazay2017} extends the PSI functionality to more than two parties, \ie, the parties jointly compute the overall intersection of all their input sets without leaking any information of set elements that are not included in the intersection.
Unfortunately, multi-party PSI only outputs the set intersection of \emph{all} input sets.
However, in our exploit depletion system we search for intersections between \emph{at least} two sets.
A possible solution to this is to implement two-party PSI protocols between each pair of parties. However, this would require a quadratic number of protocol runs in the number of parties.
Even more critically, this approach would reveal \emph{which} other party has a common vulnerability.
Instead, we use a generic \ac{mpc}-based approach for our \mpcextrust\ application that is based on \etal{Huang}'s~\cite{Huang2012} Boolean circuit for two-party PSI which we extended into a multiple parties variant.
\newline
\subsubsection{Instantiation}
\label{sec:mpc_implementation}
There are many \ac{mpc} frameworks based on secret sharing and/or garbled circuits, \eg,~\cite{MPSPDZ, SCALEMAMBA, Chaudhari2019b, Patra2020, Chaudhari2019, Bogdanov2008, Mohassel2018, Demmler2015}.
\cref{tab:mpc_frameworks} lists and compares several \ac{mpc} frameworks with malicious security.

Since untrusted actors deal with highly sensitive information, we need security against malicious parties actively manipulating the computation to either learn more information or prevent other parties from receiving the correct output.

Current \ac{mpc} frameworks that meet these requirements are MP-SPDZ~\cite{MPSPDZ} and SCALE-MAMBA~\cite{SCALEMAMBA}.
We recommend the use of~\emph{MP-SPDZ}, which implements, among other protocols, the constant-round BMR protocol~\cite{Beaver1990}, which has benefits over the multi-round protocols of SCALE-MAMBA in high-latency networks.
BMR is secure against malicious parties and a dishonest majority~(\ie, up to~$N-1$ parties can be corrupted).
If the number of computing servers is fixed to~$N=~3$~one can use ABY3~\cite{Mohassel2018}, Sharemind~\cite{Bogdanov2008}, ASTRA~\cite{Chaudhari2019}, or BLAZE~\cite{Patra2020}; if the number is fixed to~$N = 4$~parties, Trident~\cite{Chaudhari2019b} can be utilized.
The MOTION framework~\cite{MOTION} allows \ac{mpc} among any number of parties~$N$, however, it does not fulfil the full-threshold requirement of~$t=N-1$. 
Table \ref{tab:mpc_frameworks} shows an overview of the mentioned \ac{mpc} frameworks.

\begin{table}[tbp!]
\centering
\begin{tabular}{l|r|r}
\textbf{Framework} & \textbf{\# Parties $N$} & \textbf{Threshold $t$}  \\ \hline \hline
ABY$^3$~\cite{Mohassel2018}             & 3                       & 1                         \\ \hline
Sharemind~\cite{Bogdanov2008}           & 3                       & 1                         \\ \hline
ASTRA~\cite{Chaudhari2019}              & 3                       & 1                         \\ \hline
BLAZE~\cite{Patra2020}                  & 3                       & 1                          \\ \hline
Trident~\cite{Chaudhari2019b}           & 4                       & 1                             \\ \hline
MOTION~\cite{MOTION}                    & $\geq 2$                & 1 \\\hline
SCALE-MAMBA~\cite{SCALEMAMBA}           & $\geq$ 2                & $N - 1$                          \\ \hline
MP-SPDZ~\cite{MPSPDZ}                   & $\geq$ 2                & $N - 1$                         \\
\end{tabular}
\caption{Comparison of \ac{mpc} frameworks that are secure against malicious adversaries, compute on Boolean circuits and allow up to~$t$ corruptions.}
\label{tab:mpc_frameworks}
\end{table} 

\subsection{Feasibility of \mpcextrust\ implementation}

\mpcextrust\ completely relies on the security properties of the underlying multi-party computation~(\ac{mpc}) framework. 
While most \ac{mpc} frameworks are implemented for academia usage, Bosch developed \emph{Carbyne Stack} an open-source cloud stack for scalable \ac{mpc} applications~\cite{carbynestack} that is also suited for real-world usage.
As the name suggests, the long-term plan is to make this \ac{mpc} framework scalable for many participating parties. As this entire project is open-source, a group of states can use their implementation as basis \mpcextrust.

\subsection{Evaluation of the scalability of \mpcextrust}
\label{sec:mpcevaulation}

\begin{table}[tbp!]
\centering
\begin{tabular}{r||r|r|r|r}
\textbf{\# Vulnerabilities \textbackslash ~\# States} & \textbf{2} & \textbf{5} & \textbf{10} & \textbf{15} \\ \hline \hline
\textbf{100} & 2 & 14 & 62 & 146 \\ \hline
\textbf{500} & 4 & 31 & 134 & 314 \\ \hline
\textbf{1000} & 7 & 49 & 210 & 492 \\ 
\end{tabular}
\caption{Runtime in minutes of \mpcextrust{} for various numbers of maximum vulnerabilities and states.}
\label{tab:runtime_eval}
\end{table} 

In this section, we estimate the feasibility and scalability of our \mpcextrust. Since we know the complexity of our Boolean circuit, we can estimate the scalability of~\mpcextrust.

In a realistic setting of our proposed application context of arms control, we have the following parameters for our benchmarks in \cref{tab:runtime_eval}: number of parties / states $N \in \{2, 5, 10, 15\}$, maximum number of inputs~$u \in \{500, 1000, 1500\}$, and length of vulnerability identifier hashes~$\sigma = 256$ bit.
With these parameters, the size of our Boolean circuit is $\approx 4.8 \cdot 10^7$ \texttt{AND}s.

To estimate the runtime of our system, we generate a random circuit with the same number of \texttt{AND} gates and two \texttt{XOR} gates per \texttt{AND} gate.
Since \texttt{XOR} gates can be evaluated in the BMR protocol without any communication~\cite{Lindell2016}, it is less important to determine the exact number of \texttt{XOR} gates, as communication is the bottleneck of \ac{mpc}.

For malicious \ac{mpc} with a dishonest majority, as required by our adversary model presented in~\cref{sec:adversary}, we use the constant-round BMR protocol~\cite{Beaver1990} using the MASCOT protocol~\cite{Keller2016} to compute the garbled tables as implemented in the MP-SPDZ framework~\cite{MPSPDZ}.
To conduct our experiments, we use five servers, each equipped with an Intel Core i9 processor with 2.8 GHz and 128 GB DDR4-RAM.
The round-trip network latency in our simulated WAN setting is about 100~ms and the bandwidth 90~Mbit/sec.
We take the average runtime of three~executions.

The execution time of our circuit is about 31 minutes.
This is an acceptable runtime for governmental actors, as the protocol is run daily or weekly.
However, the size of the Boolean circuit and the cost of computing each \texttt{AND} gate are quadratic in the number of servers~$N$.
Therefore, our scheme will not scale for a large number of parties~$N \gg 10$.

We can improve the scalability for these scenarios by outsourcing the computation to~$n \ll N$ non-colluding servers~\cite{Kamara2011}.
Here, the~$N$ parties distribute their input to the~$n$ servers, which together run the \ac{mpc} protocol and distribute the result.
An advantage of this method is that all~$N \gg n$ parties may be malicious as long as they can trust that the~$n$ servers are not colluding.
This improves the cost of computing an \texttt{AND} gate to~$\mathcal{O}(n^2)$.

\section{Discussion}
\label{cha:discussion_and_application}

This section will discuss our approach. As the main contribution of this paper is the~\mpcextrust, the \bcextrust~prototype is not covered here, as it was discussed in \cref{cha:blockchain_limitations}. In the following, we will analyze our~\mpcextrust\ regarding the conceptual requirements RC1 - RC5~(\cref{sec:mpc_conceptual}) and the security requirements RS1 - RS3~(\cref{sec:mpc_security}) necessary to create incentives for states to participate.
\change{This section also reviews the scenarios in which \extrust\ can be of use, followed by an outlook on possible future applications~(\cref{cha:application}).
An overview of which conceptual and security requirements are fulfilled by \mpcextrust\ and \bcextrust, respectively, is provided in~\cref{tab:comparison}. The bracketed checkmarks in the table highlight requirements, that are only fulfilled if we can exclude the 51$\%$ attack against Blockchains.}

\begin{table}[t]
    \centering
    \begin{tabular}{l|c|c}
        \textbf{Requirement} & \textbf{\bcextrust} & \textbf{\mpcextrust}  \\
        \hline
        \hline
         RC1 & \reqmaybe & \reqyes \\
         \hline
         RC2 & \reqyes & \reqno \\
         \hline
         RC3 & \reqyes & \reqyes \\
         \hline
         RC4 & \reqyes & \reqyes \\
         \hline
         RC5 & \reqyes & \reqyes \\
         \hline
         \hline
         RS1 & \reqno & \reqyes \\
         \hline
         RS2 & \reqmaybe & \reqyes \\
         \hline
         RS3 & \reqyes & \reqyes \\
    \end{tabular}
    \caption{Comparison of which conceptual~RC1 - RC5 and security requirements~RS1 - RS3 are fulfilled by \bcextrust and \mpcextrust.}
    \label{tab:comparison}
\end{table}

\subsection{Conceptual Requirements}
\label{sec:mpc_conceptual}

The \mpcextrust\ architecture allows participating parties to input information about their known vulnerabilities and exploits without openly revealing sensitive information to other parties (RC1).
The output of the computation is the matching vulnerabilities and exploits between the parties (RC3).
Since the output is computed interactively between the parties, \mpcextrust\ is entirely decentralized and does not require a trusted third party~(RC5).

The solution is theoretically scalable to~$N=10$~parties (RC4). 
However, the more parties are involved in the protocol, the more inputs and data have to be exchanged between these parties, \ie, the approach has a complexity~$\mathcal{O}(N^2)$, but usually~$N \approx 10$~(cf.~\cref{sec:requirements}).
However, as explained in RC4, real time computability is not a critical requirement and longer computation times are no problem for such highly politically organized processes like arms control, which often require days or weeks for the full formal process and the involvement of all necessary stakeholders. 
In~\cref{sec:mpcevaulation}, we propose to outsource~\cite{Kamara2011} the computation to~$n \ll N$ parties, which improves the performance of \mpcextrust.
Considering the context of arms control, such a scenario is only applicable and likely if the outsourced computation is performed by neutral institutions that are not involved in the arms control measure itself, since in this way none of the parties involved need to trust that the other participants will not share information outside the protocol.
Such delegation is not uncommon for arms control measures.
An example is the Joint Comprehensive Plan of Action (JCPOA), a multilateral treaty known as the Iran nuclear deal~\cite{Maslov2020} between Iran, China, France, Russia, the United Kingdom, and Germany. 
The International Atomic Energy Agency (IAEA) manages and organizes all aspects of this treaty via independent bureaus, entrusted laboratories, UN working groups, and neutral experts for investigation field trips. 
Regardless, for practical arms control measures as our proposed depletion system, the amount of involved parties usually does not exceed a single-digit number and is often established between a small group of states.

Unfortunately, requirement RC2 is not met because intersection checks now require interaction, as the participating parties are required to exchange data.
However, exactly this property of  \extrust\ is the key to avoid local brute-force attacks, to which \bcextrust~is vulnerable~(cf.~\cref{cha:blockchain_approach}). 
Although, in the context of arms control, the minimum threshold for cooperation to which states must commit provides an incentive to join the measure, this requirement is not mandatory to practically operate \extrust. As a privacy-preserving arms control measure is more critical than the desire to independently check for intersections, we consider this a weak limitation that does not undermine the practical value of our approach, especially when considering that \extrust\ fulfils all other conceptual requirements.

\subsection{Security Requirements}
\label{sec:mpc_security}

\mpcextrust\ fulfils all three security requirements presented in~\cref{sec:security_requirements}.
A notable advantage of \ac{mpc}-based protocols is that the participating parties can only derive information from their own inputs and the outputs received, \ie, the parties do not learn more information in the \mpcextrust\ than in \extrust\ with a trusted third party that receives the inputs from all parties and outputs the intersections.
This means that no more information is revealed in the protocol transcript than an adversary would learn in the ideal world.
In contrast to \bcextrust\ from~\cref{cha:blockchain_approach}, local attacks (\eg, brute-forcing specific hash values) are not possible in \mpcextrust.
In addition, an adversary is not able to copy vulnerabilities or exploits from other parties to output an invalid intersection because the inputs are inaccessible to the other parties.
Thus, requirement RS1 is completely fulfilled.

Once the inputs are submitted in the \ac{mpc} protocol, the parties are not able to withdraw or modify them (RS2).
A situation in which all state actors jointly manipulate the protocol will never happen, since they could otherwise share their vulnerabilities in plain anyway.

False positive intersection~(RS3) results are possible with a negligible probability.
A false positive is possible if two different vulnerabilities are mapped to the same hash value.
Since we use a collision-resistant hash function, the probability of other collision scenarios is negligible.
In addition, a false positive may occur if two parties independently choose the same key identifiers.
Due to the usage of 256 bit key identifiers and a robust random generator, the probability of this situation is negligible as well.

Above all, the security and confidentiality of the assets to be shared are key incentives for establishing an arms control measure. As \mpcextrust~fulfils all security requirements, it is suitable for a real world application without discouraging states from using it.

\subsection{Further Application Scenarios}
\label{cha:application}

Beside the proposed context, \extrust\ can also be useful in other application scenarios, some of which will be discussed in the following.


At present, our approach concentrates exclusively on state actors as addressees.
However, organizations or individuals might also be interested in using such a system.
As explained in \cref{sec:externaldepletion}, \textbf{bug-bounty programs} and \textbf{vulnerability research projects} have similar goals: to reduce the spread of vulnerabilities to secure systems.
Here, using the aggregated information from the external stockpile depletion measures and integrating it into \extrust\ can increase the speed of detection of matching rediscoveries in stockpiles.
This can be achieved by using writers for selected public services or other institutions that intend to contribute to cybersecurity, which feed their hashed vulnerability identifiers into \extrust.
In such a setting, the hashing of information is as important as in \extrust's motivational scenario to prevent the material from being disseminated for malicious cyber operations.
The use of \extrust\ in a purely corporate environment is probably not possible, as organizations like Zerodium~\cite{zerodium_zerodium_2019} are primarily looking for exploits to sell.
A similar bug-bounty related approach could focus on examining discovered, potential \textbf{zero day vulnerabilities} against other submitted but not yet publicly disclosed vulnerabilities.
The history of submissions would allow submitting actors to claim their first-submitted-reward later on, once the information is disclosed.
In this way, the first finder could be paid out without the hackers having to reveal their discovery in advance.
\section{Conclusion and Future Work}
\label{cha:conclusion}
\change{Given the continuous developments in the field of cybersecurity and especially the expected advantages of using artificial intelligence measures to detect, mitigate or even defend against cyber threats, the exclusive knowledge of vulnerabilities is an essential component for state actors to stay ahead of competitors. Under the assumption, that this undermines national as well as international cybersecurity, our paper focused on the depletion of vulnerabilities and exploits that are being stockpiled by state actors.}
While the disclosure of vulnerabilities at the national level through regulatory processes is becoming more and more of an issue, cooperation on disclosure at the bilateral or multilateral level is still lacking.
We discussed that an important obstacle to such measures is the comprehensible restraint of states to give up their accumulated intelligence information in order to compare stockpiles and unnecessarily reveal unique exploits or other secret assets. \\
\change{To} develop a technical measure in such a zero-trust scenario, we identified structural as well as IT security requirements for the detection of intersections in different exploit stockpiles.
Based on these, we discussed and designed
\begin{enumerate*}[label=(\roman*)]
	\item a novel identification scheme for vulnerabilities and exploits and
	\item an external, privacy-preserving exploit depletion system named \extrust.
\end{enumerate*} We have identified the requirements for this depletion system for zero-trust relationships 
and shown that the technical security requirements could hamper the political incentives for states to cooperate. 
We have illustrated this challenge by developing a prototype for a depletion system based on a Blockchain. 
The presented \mpcextrust\ system handles this dualism by focusing on the IT security of a depletion system while fulfilling most of the conceptual requirements.  
It stores the detected intersections, while the submitted vulnerabilities are protected by the \ac{mpc} protocol and thus remain hidden from all involved actors. 
However, one limitation of this approach is that it is vulnerable to \change{secret agreements} by multiple actors, as they could add vulnerabilities and remove them from the intersection -- an edge case that is not an option in the proposed arms control context. We also argued that \mpcextrust\ is currently not able to fulfil all conceptual requirements, as participating states need to explicitly cooperate and share obfuscated information, which could be a disadvantage regarding its implementation.
Nevertheless, we have shown that the strength of the \ac{mpc} protocol lies in the fact that an adversary cannot obtain more information from the joint computation than if a trusted third party were to compute the intersections.
The \extrust\ system uses a novel exploit identifier and discussed how this identifier could be improved in different scenarios to address the trade-off between the uniqueness and ambiguity of the properties. 
We believe that this provides a secure measure which fulfils the state's need for secrecy and yet at the same time can contribute to the reduction of vulnerability stockpiles \change{to} foster the public IT security through the disclosure of vulnerabilities. We discussed further application scenarios beyond the specific context of cyber arms control with different parties comparing their vulnerability stockpiles. 
We demonstrated that such an approach could be facilitated for external depletion measures such as bug-bounty programs.
Such measures could potentially be extended so that even private actors could contribute to the internal exploit stockpile depletion process by adding external information about the depletion into \extrust. \\ 
 
\change{As discussed}, further evaluation and study of our concept is recommended, in particular in terms of the definition of the identifier. 
We discussed that a current limitation of the identifier is the necessity to find a sweet spot in the accuracy regarding the description of a security vulnerability that prevents duplicate descriptions of the same identifier while avoiding an unnecessary and potentially problematic generalization. Future work should analyze the relationship between the uniqueness and ambiguity of the characteristics of the identifier, the size of the identifier space, and -- on a practical level -- whether security experts independently create matching identifiers for the same vulnerability. Further work should focus on the possibility, the role and the security requirements of a trusted third party like the UN to calculate stockpile intersections, \change{to} circumvent the current necessity of cooperation between potentially opposing state actors.
In addition, it would be interesting to implement \extrust ~as an actual measure between state parties to monitor its real world usage, its perception of the systems security and usability by the participating states as well its impact on their vulnerability disclosure considerations. \\

Due to the high political relevance of our proposal, we hope that this approach can be an inspiration to computer science and engineering to reflect on the ethical responsibility for the domain of cyberspace and its peaceful development and that future interdisciplinary work in this area will bring together researchers from privacy, IT security, and peace and conflict research.

\ifanonymous
\else
\section*{Acknowledgements}
This research work has been funded
by the Deutsche Forschungsgemeinschaft (DFG, German Research Foundation) -- SFB 1119 (CROSSING) -- 236615297.
It was co-funded by GRK~2050 Privacy \& Trust/251805230,
the German Federal Ministry of Education and Research and the Hessian Ministry of Higher Education, Research, Science and the Arts within their joint support of the National Research Center for Applied Cybersecurity ATHENE, as well as
the European Research Council~(ERC) under the~European Union's Horizon~2020 research and innovation program~(grant agreement No.~850990 PSOTI).

\fi

\bibliographystyle{IEEEtran}
\bibliography{bibliography.bib}

\iffullversion
\section*{Annex}
\label{cha:annex}

\subsection{Required properties for a machine-readable vulnerability identifier}
\label{sec:vulnerabilityIdentifier}

In order to automatically compare vulnerabilities, our approach requires a vulnerability identifier~$\mathit{id}(i^p_v)$ for a given vulnerability~$v$, which must be unique among all parties~$p \in \mathbb{P}$ sharing~$v$'s~information~$i_v$-. Hence, the equation~$\mathit{id}(i^{p}_v) = \mathit{id}(i^{p'}_v)$ should hold for all~$v \in \mathbb{V}$ and~$p, p' \in \mathbb{P}$, where~$i^p_v$ and~$i^{p'}_v$ are the vulnerability~$v$'s information of the respective party.

\subsection{Ambiguous vulnerability identifier}
\label{sec:ambiguousVulnerabilityIdentifier}

The vulnerability identifier description is ambiguous, if it is possible to describe two different vulnerabilities~$v_1, v_2 \in \mathbb{V}$, where~$v_1 \not= v_2$ with the same identifier, \ie,~$\mathit{id}(i^{p}_{v_1}) = \mathit{id}(i^{p}_{v_2})$, or to use two different identifiers for the same vulnerability~$v \in \mathbb{V}$, \ie,~$\mathit{id}(i^{p}_v) \not= \mathit{id}(i^{p'}_v)$.

\subsection{Approximation of the vulnerability identifier space}
\label{sec:vulnerabilityIdentifierSpace}

Given the number of \ac{cwe} classes, the size of the \ac{cpe} directory and space of possible function names~$\mathit{FN}$ the vulnerability identifier space can approximated to~$|id| = |\mathit{CPE}|*|\mathit{CWE}|*|\mathit{FN}|$, with~$|\mathit{CPE}| \approx 2^{19}$, $|\mathit{CWE}| \approx 2^9$, and~$|\mathit{FN}| \approx 2^{53}$~(using the NLTK English word corpus) resulting in~$|\mathit{id}| \approx 2^{81}$.

\subsection{The \texttt{checkIntersections} transaction of \bcextrust}
\label{sec:checkIntersections}

\begin{lstlisting}[caption={The function that checks for matching items in the \bcextrust~ledger.},label=lst:checkIntersections]
/**
 * Sample transaction processor function.
 * @param {de.peasec.vulnerability.CheckCollisions} tx Sample transaction instance
 * @transaction
 */
async function checkCollisions(tx) {
	const assetRegistry = await getAssetRegistry('de.peasec.vulnerability.Exploit');
	const allExploits = await assetRegistry.getAll();
	const collidingExploits = {};

	for (const exploit of allExploits) {
		const id = exploit.id;
		const hash = exploit.vulnerabilityIdentifierHash;

		if (!(hash in collidingExploits)) {
			collidingExploits[hash] = [exploit];
		} else {
			collidingExploits[hash].push(exploit);
		}
	}

	for (const property in collidingExploits) {
		if (collidingExploits.hasOwnProperty(property) && collidingExploits[property].length >= 2) {
			exploits = collidingExploits[property];

			let event = getFactory().newEvent('de.peasec.vulnerability', 'CollisionFound');
			event.affectedExploitIds = exploits;
			emit(event);
		}
	}

}
\end{lstlisting}

\subsection{PSI-variant Boolean circuit for multiple parties}
\label{sec:psiBooleanCirciutVariant}

Our circuit extends \etal{Huang}'s~\cite{Huang2012} Boolean circuit for two-party PSI of complexity~$\mathcal{O}(u \log u)$ into a PSI-variant circuit for multiple parties with special-purpose filter options for matching items~(cf.~\cref{sec:circuitGeneralization}).
In the first step, each party~$i$ locally sorts (\ie, outside the circuit) its set of triples~$X_i = \{x_{i, 1}, \dots, x_{i, u}\}$ s.t.~$v_{i, j} < v_{i, l}$ if~$j < l$ and inputs this into the circuit.
The first task of the circuit is to verify that the input set of each party is correctly sorted.
This is a linear sweep through the input set of each party with~$u$ comparison circuits for the respective~$v_{i, j}$-values and has total size of~$\mathcal{O}(Nu\sigma)$ \texttt{AND}s~\cite{Kolesnikov2009}.
We open a flag~$l_i$ for each party~$i$ to all parties by using reactive \ac{mpc}, that indicates if~$i$'s inputs were correctly sorted.
If one of the parties cheats~($l_i = 1$), the protocol aborts and the cheating parties are removed from future computations.

The next part of the circuit is similar to that of \etal{Huang}~\cite{Huang2012}: the single sorted sets are oblivious merged to one large sorted set.
For this purpose, we span a binary tree of sorted lists, where each node on depth~$i$ implements a bitonic merger~\cite{Huang2012} for~$2^iu$ inputs, which has a complexity of~$\mathcal{O}( 2^{i}u\sigma \log(2^iu))$ \texttt{AND}s.
A bitonic merger takes two sorted sets as input and outputs the merged sorted set.
In total, we have~$N-1$ of these bitonic mergers and the binary tree has a depth of~$\log_2(N)$.
This results in an upper limit of~$\sim2N^2 u \sigma \log(Nu)$ \texttt{AND}s.
In our special case, all triples~$x_{i, j}$ are now sorted according to the values~$v_{i, j}$ of $x_{i, j}$, so that matching vulnerabilities lie directly next to each other.
Since the triple can no longer be assigned to a specific party, we denote the outcome of this subcircuit with~$x_1, \dots, x_{uN}$.

\begin{figure*}[htbp!]
    \centering
    \includegraphics[scale=0.18]{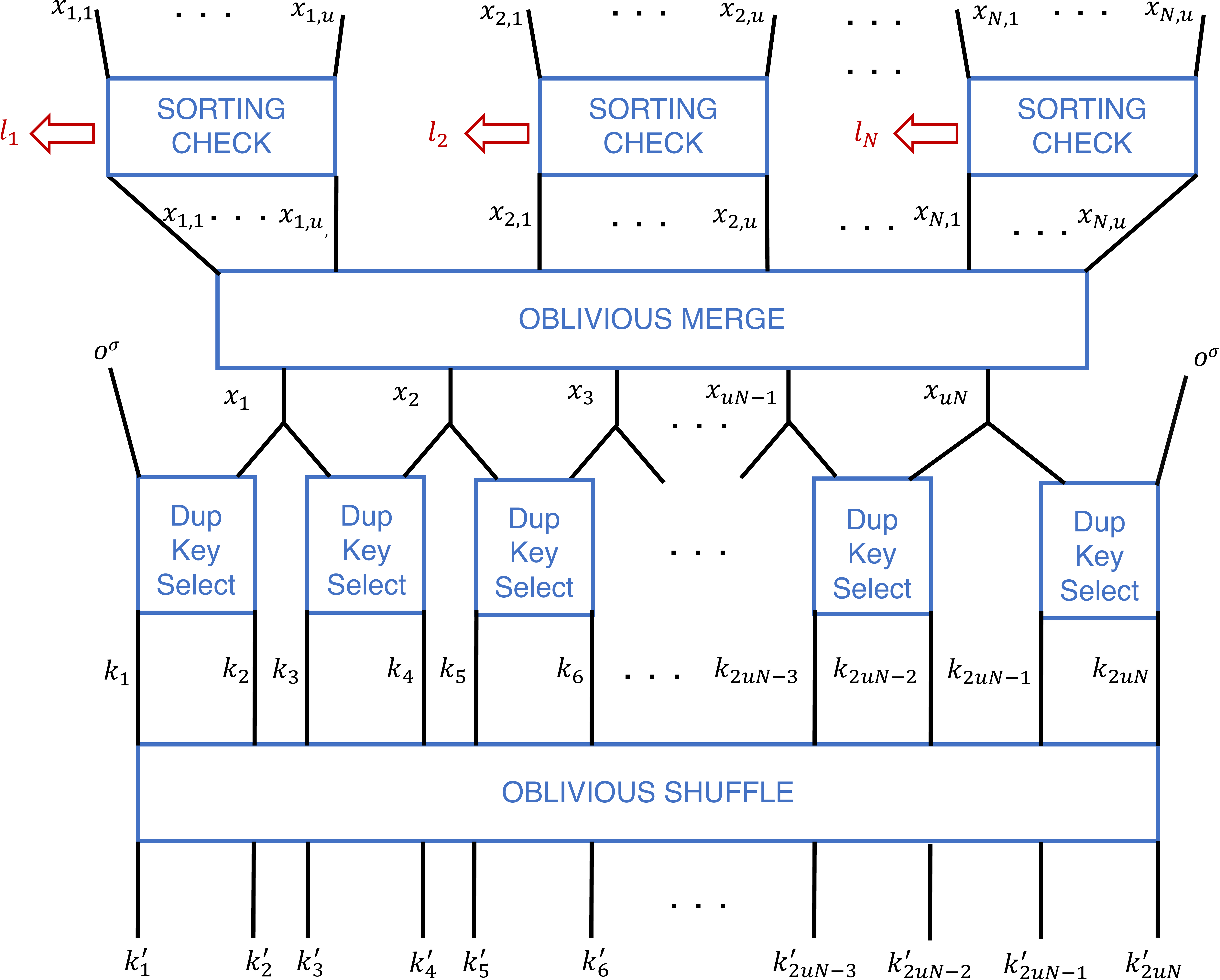}
    \caption{Boolean Circuit design output identifier keys that are dependent if their respective vulnerability identifier occurs at least twice. The complexity of the circuit is~$\sim2Nu + 2N^2 u \sigma \log(Nu)$ \texttt{AND}s for number of parties~$N$, number of inputs per party~$u$ and security parameter~$\sigma$. The design is based on \etal{Huang}'s Boolean circuit for computing the set intersection between two parties~\cite{Huang2012}. The red values~$l_1, \dots, l_n$ are opened and verified before the oblivious merge block is executed.}
    \label{fig:mpccircuit}
\end{figure*}

Afterwards, neighboring entries are compared with a \emph{DupKeySelect} block.
A DupKeySelect block takes as input two neighboring triples~$(v_{i}, k_{{i}}^0, k_{{i}}^1)$ and~$(v_{i+1}, k_{{i+1}}^0, k_{{i+1}}^1)$, and outputs~$(k_{{i}}^1, k_{{i+1}}^1)$ if~$v_{i} = v_{i+1}$ or~$(k_{{i}}^0, k_{{i+1}}^0)$ otherwise.
We compare the first and last value with a zero string~$0^\sigma$ to avoid leaking information about the frequency of occurrence of a key.
The corresponding circuit a has size of~$\sim Nu\sigma$ \texttt{AND}s.

The final step is to shuffle the output keys~$k_1, \cdots, k_{2uN}$. 
This circuit has a complexity of~$\mathcal{O}(Nu\sigma \log (Nu))$ \texttt{AND}s~\cite{Waksman68}.
The resulting keys~$k_1^\prime, \dots, k_{2uN}^\prime$ are opened to all parties.
Finally, the parties can identify matching vulnerabilities by checking which of their input keys occur in the output.
Overall, the complexity of the circuit is clearly dominated by the oblivious merge part of size~$\sim 2N^2 u \sigma \log(Nu)$ \texttt{AND}s.

During the lifetime of \extrust, the parties cannot update the hash values~$v_{i,j}$ from~$x_{i,j}$, but they use fresh identifier keys~$k_{i, j}^0$ and~$k_{i, j}^1$ in each protocol run.

\subsubsection{Generalization of \mpcextrust}
\label{sec:circuitGeneralization}

The above described circuit implements a variant of the private multi-party intersection, where an intersection is found when at least two parties share an element.
However, this circuit can easily be adapted to other variants of the functionality relevant for arms control.
We provide two examples below and note that adapting to other variants is a simple circuit design task.

\paragraph{At least~$m$ parties} 
In some cases, it may be useful to know whether elements are shared by at least~$m$ parties~(say $m=3$) in order to decide whether a vulnerability or exploit can be made public.
This can be easily achieved by replacing the \emph{DupKeySelect} block in~\cref{fig:mpccircuit} with an \emph{$m$-DupKeySelect} block, which outputs the~1~keys if~$m$ neighboring elements are equal.
Such an~\emph{$m$-DupKeySelect} block has a complexity of~$\mathcal{O}(m\sigma)$ \texttt{AND}s so that the total complexity of this subcircuit is~$\mathcal{O}(mNu\sigma)$ \texttt{AND}s which is still negligible compared to the rest of the circuit.

\paragraph{At least~$z$ fixed parties and~$m$ other parties}
In this scenario, the parties aim to find intersections that~$z$ fixed parties know about~(\eg, China and USA with~$z=2$) and at least~$m$~other parties.
This can be achieved by adding a fixed identifier~$t$ to the input tuples of the parties~(\eg, set to 0 for China, 1 for the USA, 2 for all others).
The \emph{DupKeySelect} block in~\cref{fig:mpccircuit} is then replaced by a \emph{Programmable-$(m+z)$-DupKeySelect} block, which receives a programming bit~$p$ as additional input, indicating whether a duplicate check is valid or not.
More specifically, if~$p=1$, the block will output the result of an \emph{$(m+z)$-DupKeySelect} block, and otherwise, if~$p=0$, the block will never output an intersection.
The programming bit for these blocks is indicated by a~\emph{$z$-Filter} block, which takes the parties' identifier~$t$ and checks if the~$z$ fixed parties are part of the potential intersection~(\eg, $t \in \{0, 1\}$).

The~\emph{$z$-Filter} block takes as input~$z+m$ identifiers~$t_1, \dots, t_{z+m}$, has a set of fixed identifiers~$T_1, \dots, T_z$, outputs one single bit~$p$ and works as follows:
We check for all fixed identifiers~$T_i$ if an input identifier~$t_j$ exists where~$T_i = t_j$ holds.
We do this by comparing~$T_i$ to all input identifiers and using OR-gates to fold the result to a single bit.
The total complexity of this step for all fixed identifiers is~$\mathcal{O}(z\omega(m+z))$ \texttt{AND}s, where~$\omega=\log(z+1)$ is the bit-length of the state identifiers.
Afterwards, we use~$z-1$ \texttt{AND} gates to fold the~$z$ resulting bits to one bit~$p$, which results in a total complexity of~$\mathcal{O}(z\omega(m+z))$ \texttt{AND}s.

The \emph{Programmable-$(m+z)$-DupKeySelect} block takes~$z+m$~quadruples~$x_i = (v_i, k_i^0, k_i^1, t_i)$ as input and outputs~$z+m$ keys~$k_1^{0/1}, \dots, k_{z+m}^{0/1}$.
It consists of a~$(z+m)\sigma$-bit multiplexer (one bit for each output bit), where its first input consists of the input keys~$k_0^0, \dots, k_{z+m}^0$, the second input is the output of the~\emph{$(m+z)$-DupKeySelect} block, and the programming bit is the output of the~\emph{$z$-Filter} block~$p$.
The multiplexer circuit has a size of~$\mathcal{O}(\sigma(z+m))$ \texttt{AND}s.
Overall, the complexity of the~\emph{Programmable-$(m+z)$-DupKeySelect} block is dominated by the size of the~\emph{$(m+z)$-DupKeySelect} block of size~$\mathcal{O}(mNu\sigma)$ \texttt{AND}s which is negligible compared to the rest of the circuit.

\fi

\end{document}